\documentclass[12pt, titlepage, reqno]{article}

\usepackage[cmex10]{amsmath}
\usepackage{libertine}

 \usepackage{bm}
 \usepackage[mathscr]{eucal}
 \usepackage{enumerate}
 
 \usepackage{amssymb,amsmath,amsthm,amscd}
\usepackage{mathrsfs}

\usepackage[T1]{fontenc}
\usepackage[caption = false]{subfig}

\usepackage{float}

\usepackage{algpseudocode}
\usepackage{algorithm,algpseudocode}
\usepackage{graphicx}

\usepackage{tabularx}

\usepackage{multirow}

\usepackage{hyperref}
\hypersetup{
    colorlinks,%
    citecolor=black,%
    filecolor=black,%
    linkcolor=black,%
    urlcolor=black,
}  

\usepackage{amssymb}

\renewcommand{\P}{{P}}

\newcommand{\tends}[2]{\displaystyle\mathop{\longrightarrow}_{#1\rightarrow#2}}

\makeatletter
\def\captionof#1#2{{\def\@captype{#1}#2}}
\makeatother

\begin{document}

 \begin{titlepage}

 \begin{center}

\textbf{Understanding the Probabilistic Latent\\ Component Analysis Framework}

\vspace{10ex}

Dorian Cazau$^{a}$ \footnote{Corresponding author e-mail: dorian.cazau@ensta-bretagne.fr}, Gregory Nuel$^{b}$

\vspace{1ex}

\begin{flushleft}

\small{$^a$ ENSTA Bretagne - Lab-STICC (UMR CNRS 6285), Universit\'e Europ\'eenne de Bretagne, 2 rue Fran\c{c}ois Verny, 29806 Brest Cedex 09, France} \\

\small{$^b$ Laboratoire de Math\'ematiques Appliqu\'ees (MAP5) UMR CNRS 8145, Universit\'e Paris Descartes, Paris, France} 

\end{flushleft}

 \end{center}

 \end{titlepage}

\begin{abstract}
Probabilistic Component Latent Analysis (PLCA) is a statistical modeling method for feature extraction from non-negative data. It has been fruitfully applied to various research fields of information retrieval. However, the EM-solved optimization problem coming with the parameter estimation of PLCA-based models has never been properly posed and justified. We then propose in this short paper to re-define the theoretical framework of this problem, with the motivation of making it clearer to understand, and more admissible for further developments of PLCA-based computational systems.

\end{abstract}

\addtocounter{page}{2}

\section{Introduction}

\subsection{Background}

Methods of Probabilistic Latent Analysis arise from the aspect model \cite{Hofmann1999b,Saul1997}, which in turn belongs to the family of statistical mixture models \cite{McLachlan1988}. Such methods provide a solid statistical foundation, as involving the likelihood principle as well as a proper generative model of the data. This implies in particular that standard techniques from statistics can be applied for questions like model fitting, model combination, and complexity control. First applications of these methods were made on semantic indexing of text corpus, with the Probabilistic Latent Semantic Indexing framework \cite{Hofmann1999} developed upon the Latent Semantic Analysis. The factor representation obtained by this method allows to deal with polysemous words and to explicitly distinguish between different meanings and different types of word usage. Within this framework, Probabilistic Latent Component Analysis (PLCA) has then been developed as a general method for feature extraction from non-negative data, with pioneer applications to audio \cite{Smaragdis2006} and image \cite{Smaragdis2008}. Following studies in audio research have in particular dealt with the tasks of multi-pitch estimation \cite{Benetos2011d,Fuentes2013,Arora2013,Fuentes2014,Sung2014,Arora2015}, sound source separation \cite{Ganseman2012,Mohammadiha2013b}, instrument identification \cite{Arora2014}, melody extraction \cite{Han2011,Fuentes2012b}, temporal music structure \cite{Weiss2011} and speech processing \cite{Kawaguchi2012,King2012}.

\subsection{General Formulation}

Aspect model \cite{Hofmann1999b,Saul1997} is a latent variable model for general co-occurrence data which associates an unobserved class variable $z \in \boldsymbol{Z}=\{z_1, \cdots, z_K \}$ with each observation, i.e. with each occurrence of an acoustic event $e \in \boldsymbol{E}=\{e_1, \cdots, e_M \}$ belonging to a different group of events $g \in \boldsymbol{G}=\{g_1, \cdots, g_N \}$. The generative model associated with this formalism is defined as follows

\begin{itemize}
\item select an event group $g$ with probability $\P(g)$,
\item pick a latent class $z$ with probability $\P(z | g)$,
\item generate an acoustic event $e$ with probability $\P(e | z)$.
\end{itemize}

The mathematical expression of this process takes the form of a a joint probability model, where one has to sum over the possible choices of $z$ which could have generated the observed pair $(e,g)$, i.e. 

\begin{equation}\label{General_PLCA}
\P(e,g) = \P(g)\sum\limits_{z\in\boldsymbol{Z}}  \P(e | z) \P(z | g)
\end{equation}

Such a generative process follows two independence assumptions \cite{McLachlan1988}: 

\begin{enumerate}
\item Observation pairs $(e,g)$ are assumed to be generated independently; 
\item The conditional independence assumption is made that conditioned on the latent class $z$, acoustic events $e$ are generated independently of the specific event group $g$.
\end{enumerate}

For speech modeling with the PLSI method \cite{Hofmann1999}, the acoustic event $e$ corresponds to words and the group of events $g$ corresponds to documents. For music modeling with the PLCA method \cite{Smaragdis2006}, the acoustic event e corresponds to frequencies and the group of events $g$ corresponds to time frames.

\subsection{Fitting PLCA model}

The classical data available for fitting PLCA model is an empirical distribution $\pi(e,g)$ over the bi-dimensional space of events and groups. This distribution can be produced directly from observed data, which can be a corpus of words gathered in different documents as in the PLSI method \cite{Hofmann1999}, a spectrogram of a musical excerpt as used in most audio applications of the PLCA method \cite{Smaragdis2006,Mysore2010b,Fuentes2013,Benetos2013}, or any various frequency table. Fitting a PLCA model on such data consists in choosing $\P(g)$, $\P(z|g)$, and $\P(e|z)$ such as $\P(e,g)$ ``approximates'' $\pi(e,g)$ in a sense that is seldom (or never) specified in PLCA literature. The usual step is then to explain that this (undefined) problem can be solved using an EM-like algorithm, based on the original algorithm developed by \cite{Dempster1977}. In the next section, we will: 1) define formally the optimization problem we are trying to solve; 2) explain why and how the EM-algorithm applies to this problem.

\section{Theoretical framework}

\subsection{Optimization problem}

Stating that our PLCA model should be such as $\P(e,g)$ ``approximates'' $\pi(e,g)$ is clearly not sufficient to define our optimization problem. Since we try to approximate distribution  $\pi(e,g)$  with $\P(e,g)$, a ``natural'' approach consisting in minimizing a distribution distance between both distributions. In this context, a very popular choice is to use the Kullback-Leibler divergence \cite{Kullback1951} (this divergence is also called the relative entropy) between $\pi(e,g)$ and $\P(e,g)$:

\begin{equation}
\mathrm{KLD}(\pi| \P)= \sum\limits_{e\in\boldsymbol{E}} \sum\limits_{g\in\boldsymbol{G}}
\pi(e,g) \log \frac{\pi(e,g)}{\P(e,g)}
\end{equation}
which we would like to minimize.

By using Bayes formula and the fact that $\pi(e,g)$ does not depend on PLCA parameters, we immediately obtain that 
$\P(g)=\pi(g)=\sum_e \pi(e,g)$, and that $\P(e | z)$ and $\P(z | g)$ should be chosen such as minimizing the following objective function:
\begin{equation}
\text{fobj}(\P)=
- \sum\limits_{e,g} \pi(e,g)
 \log \left\{\sum\limits_{z}  \P(e | z) \P(z | g)\right\}.
\end{equation}

\subsection{Likelihood}

Let's define $(e_1,g_1),\ldots,(e_N,g_N)$ a $N$-sample drawn from $\pi(e,g)$, the law of large numbers gives:
\begin{equation}\label{eq:likelihood}
\frac{1}{N} \sum_{i=1}^N \log \left\{\sum_{z_i} \P(e_i| z_i)\P(z_i |g_i) \right\} 
\tends{N}{\infty} -\text{fobj}(\P).
\end{equation}
Hence, our optimization problem can be interpreted as maximizing the log-likelihood of the latent class model $\P(e | g)=\sum _{z} \P(e| z)\P(z |g)$ using an infinite sample drawn from $\pi(e,g)$.

\subsection{EM-based estimation}

With a finite sample of size $N$, maximizing the lefthand likelihood in Eq.~\ref{eq:likelihood} can be achieved iteratively by a direct application of the original EM algorithm \cite{Dempster1977}. Since $z_1,\ldots,z_N$ are latent variables in our model, the E-step of the algorithm consists in computing the following expected conditional likelihood:
\begin{equation}\label{eq:QN}
Q_N(\P | \P_\text{old})=\frac{1}{N} \sum_{i=1}^N \sum_{z_i } \P_\text{old}(z_i | e_i,g_i) \log \P(e_i, z| g_i)
\end{equation}
where $\P_\text{old}$ represents the values of PLCA parameters from the previous iteration:
\begin{equation}
\P_\text{old}(z | e,g) = \frac{\P_\text{old}(e|z)\P_\text{old}(z | g)}
{\sum_{z'} \P_\text{old}(e|z')\P_\text{old}(z' | g)}.
\end{equation}

If $N$ tends to infinity, $Q_N(\P | \P_\text{old})$ tends to
\begin{equation}
Q(\P | \P_\text{old})=\sum\limits_{e,g} 
\pi(e,g) \sum_{z} \P_\text{old}(z | e,g) \log \P(e, z| g).
\end{equation}
The M-step of the algorithm consists now in maximizing $Q(\P | \P_\text{old})$. For that purpose, we just need to find the zero of the gradient, i.e. $\forall$ $(e,z) \in \boldsymbol{E}  \times \boldsymbol{Z} $ and $\forall$ $(z,g) \in \boldsymbol{Z}  \times \boldsymbol{G}$, we want:
\begin{equation}
\left\{
\begin{array}{c}
\displaystyle\frac{\partial Q(\P | \P_\text{old})}{\partial \P(e|z)} = \frac{1}{\P(e|z)} \sum_g \pi(e,g) \P_\text{old}(z | e,g)=0\\
\displaystyle\frac{\partial Q(\P | \P_\text{old})}{\partial \P(z|g)} = \frac{1}{\P(z|g)} \sum_e \pi(e,g) \P_\text{old}(z | e,g)=0
\end{array}
\right.
.
\end{equation}
By combining these equations with the stochastic constraints $\sum_e \P(e|z) = \sum_z \P(z|g)=1$ we get immediately:
\begin{equation}
\left\{
\begin{array}{l}
\P(e|z)=\displaystyle\frac{\sum_g \pi(e,g) \P_\text{old}(z | e,g)}
{\sum_{e'}\sum_g \pi(e',g) \P_\text{old}(z | e',g)}\\
\P(z|g)=\displaystyle \frac{ \sum_e \pi(e,g) \P_\text{old}(z | e,g)}
{\sum_{z'} \sum_e \pi(e,g) \P_\text{old}(z' | e,g)}
\end{array}
\right.
.
\end{equation}

\section{Conclusion $\&$ Perspectives}

In this short paper, we have provided the necessary mathematical background to understand what problem is usually solved when fitting PLCA models to observed data, and how and why it is connected to the EM algorithm. If this innovation is in itself intellectually satisfying, one could claim that it provides nothing really useful since the resulting formulas are unchanged by this approach. However, this framework better justifies recent extensions of PLCA models towards Hidden Markov Models in order to develop joint modeling of spectral structures and temporal dynamics \cite{Mysore2010b,Benetos2013}. By connecting PLCA estimation explicitly to the standard EM algorithm, the clarified theoretical background to use extension of the EM algorithm such as Generalized EM with Newton-Raphson steps \cite{McLachlan2008}. For example, this might allow to introduce multinomial or Poisson distributed components in PLCA models.

%It also opens new perspectives of modeling development, with for example thxe use of discrete distributions (e.g. Poisson, multinomial) to constrain PLCA parameter estimation, which usually requires to switch from EM to Generalized-EM algorithm (e.g. EM step combined with a Newton-Raphson step).

%In this paper, we have provided the mathematical proofs justifying the use of the EM-algorithm to estimate the parameters of PLCA models.

%\bibliography{References_Biblio}
%\bibliographystyle{IEEEtran}


\end{document}